\documentclass[5p, sort&compress]{elsarticle}

\usepackage{amssymb,amsmath, amsfonts}
\usepackage{graphicx}
\usepackage{hyperref}
\usepackage[mathlines]{lineno}
\linenumbers

\def\shrinkfactor{1}

\usepackage[usenames,dvipsnames]{color}
\usepackage[utf8]{inputenc}
\usepackage{ulem}

\newcommand{\add}[1]{\textcolor{blue}{#1}}

\begin{document}

\title{Effective dynamics for chaos synchronization in networks with time-varying topology}

\author{R.F.~Pereira}
\ead{pereira.rf@gmail.com}

\author{R.M.~Szmoski}
\ead{rmszmoski@gmail.com}

\author{S.E.de S.~Pinto\corref{cor}}
\ead{desouzapinto@pq.cnpq.br}
\ead[url]{http://www.fisica.uepg.br/grupos/dnlsc}

\cortext[cor]{Corresponding author at: Departamento de F\'isica, Universidade Estadual de Ponta Grossa, 84030-900, Ponta Grossa, PR, Brasil. {\it E-mail}: desouzapinto@pq.cnpq.br. {\it Phone}: +55 42 3220 3044  {\it Fax}: +55 42 3220 3042.}
\address{Departamento de F\'isica, Universidade Estadual de Ponta Grossa, 84030-900, Ponta Grossa, PR, Brasil}
 
\date{\today}

\begin{abstract}
A coupled map lattice whose topology changes at each time step is studied. We show that the transversal dynamics of the synchronization manifold can be analyzed by the introduction of effective dynamical quantities. These quantities are defined as weighted averages over all possible topologies. We demonstrate that an ensemble of short time observations can be used to predict the long-term behavior of the lattice. Finally, we point out that it is possible to obtain a lattice with constant topology in which the dynamical behavior is asymptotically identical to one of the time-varying topology.
\end{abstract}

\begin{keyword}
coupled map lattice \sep time-varying coupling \sep chaos synchronization \sep effective dynamics
\end{keyword}

\maketitle

\section{Introduction}
In the physical world there are sets whose elements interact with each other. Neurons \cite{Science.v326.p1419.y2009}, individuals \cite{Nature.v446.p664.y2007}, Josephson junctions \cite{Science.v293.p843.y2001}, and computers \cite{Science.v293.p2031.y2001} are examples of elements that compose these sets called networks. Some properties of a network derive from its topology, which can be different in each of the abovementioned situations, namely: regular, random, small-world, and scale-free \cite{Nature.v410.p268.y2001}. Although the structure in each network is different, a common feature merges in these mathematical systems: the capability of its elements to synchronize. Aiming to improve the understanding on systems that present interactions between their elements, in this letter we dealt with networks whose topology varies with time \cite{Nature.v406.p845.y2000, Nature.v446.p664.y2007, Science.v325.p414.y2009, Nature.v439.p462.y2006, PhysicaD.v195.p188.y2004, PRE.v66.p016209.y2002, PRE.v78.p066209.y2008}. Based on the synchronization capability we present exact results in order to predict the long-term behavior of these networks.

 The main idea of this work is the statement of effective quantities for the analysis of the transversal dynamics to the synchronization mainifold $\mathcal{S}$, from which, for instance, the entire Lyapunov spectrum can be calculated. An important point here is that these quantities allow the construction of a lattice with an static coupling which, for synchronization purposes, exhibits behavior consistent with the networks with time-varying topology. We say that such {\it effective coupling} represents the {\it effective dynamics} of the system. We hypothesized that each realization of the original time-varying network, for a long enough observation period, can be replaced by an effective system. Such a system is obtained by calculating a weighted average over the set of possible states.

The manuscript has six main sections including the introduction. Firstly, we shall present  the class of coupled map lattice, with time-varying topology, that is the subject of this work. The next section outlines the mathematical tools for calculating the effective quantities. This will be followed by the derivation of the expression for such quantities and  the construction of the effective system. The next section will present  a application of our results in a coupled map lattice whose coupling between the elements is semirandomly determined: the coupling probability between two elements in a given instant depends algebraically on their distance in the lattice. This coupling form was studied in Refs. \cite{PRL.v74.p3297.y1995, IJBC.v17.p2257.y2007} as a model of a network of neurons. Our conclusions are left to the last section.

\section{\label{SEC.cml}Coupled map lattice}

We examined a network called {\it coupled map lattice} (CML) \cite{CHAZOTTES.JR.dynamics.of.coupled.map.lattices.and.of.related.spatially.extended.systems.y2005}, which is defined in the following way: let  $\mathbf{g}:\omega\rightarrow\omega$ be an axiom-A\footnote{A map that satisfies the axiom-A is hyperbolic and mixing \cite{RMP.v57.p617.y1985}.} $d-$dimensional map defined in $\omega\subset \mathbb{R}^d$. Let $\mathbf{F}:\Omega\rightarrow \Omega$ be a lattice with $N$ coupled maps in the form $\mathbf{g}$, with $\Omega=\omega^N$. This lattice reads
\begin{equation}\label{cml}
  \mathbf{y}_{n+1}=\mathbf{F}(\mathbf{y}_n,n)=\mathbf{G}_n\mathbf{f}(\mathbf{y}_n),
\end{equation}
in which $\mathbf{y}_n \circeq \left[ \begin{array}{ccc} \mathbf{x}_n^{(0)} & \cdots & \mathbf{x}_n^{(N-1)}\\ \end{array} \right]^T$, $\mathbf{x}_n^{(m)}$ defines the state of the $m$-th site, $\mathbf{f}^{(m)}(\mathbf{y}_n) = \mathbf{g}(\mathbf{x}_n^{(m)})$, and $\mathbf{G}_n$ is a $dN\nobreak\times\nobreak dN$ matrix that depends on discrete time $n$, in an explicit way. This is the {\it coupling matrix}.

\subsection{Topology}
We shall restrict the lattice to the linear couplings, as we can see in the equation (\ref{cml}), and consider the lattice topology varying stochastically with time. However, the following conditions must be observed:
\begin{enumerate}[(i)]
	\item the lattice is a periodic one;
	\item the topology at each time interval is invariant under translation\add{s} in the lattice.
\end{enumerate}
Since there are constraints about the topology, one can call this form of {\it semirandom} coupling. The above conditions imply $\mathbf{G}_n$ is a circulant matrix. Due to periodicity imposed to the topology, there is no reason to define a preferential direction (say, to the left or to the right) and we also require the topology to be a symmetric one. These features do not depend on the probability rule that defines each possible coupling matrix. The condition that the phase space $\Omega$ is a direct product of $N$ subspaces $\omega$ is satisfied if, and only if,
\begin{equation}\label{conditionsA}
  \left[\mathbf{G}_n \right]_{qw}\geq 0, \qquad \sum_{m=0}^{dN-1}\left[\mathbf{G}_n \right]_{qm}=1, \qquad \left( \forall q,w\right).
\end{equation} 
Herein, $q,w=0,1,\cdots,dN-1$. Thus, $\mathbf{G}_n$ is also a stochastic matrix. We imposed also that the system (\ref{cml}) represents the coupling (direct or indirect) between all the $N$ maps. That is, the state of an arbitrary site eventually influences all other elements in the lattice. If not, the system (\ref{cml}) can be splited in two uncoupled subsystems. In this case, our results are applied separately to each subsystem. Mathematically, this means that  the condition $0 < [\mathbf{G}_k]_{0w}<1$ must be satisfied, for at least one $\mathbf{G}_k$, for each $w$, in a such way that the infinite product of matrices $\mathbf{G}_k$ typically results in a strictly positive matrix \cite{PPWolfowitz, PPHajnal}. For instance, if each site is always coupled with its first neighbors, this condition is always satisfied. Here, we consider that the network topology uniquely defines the coupling matrix $\mathbf{G}_n$, {\it i.e.} $\mathbf{G}_n = \mathbf{G}_n(\mathbf{T}_n)$, in which the {\it topology matrix} $\mathbf{T}_n$ describes the connectivity between sites.

If we suppose that the rules that specify the coupling probabilities are well-defined and do not change over time, we have
\begin{eqnarray*}
  [\mathbf{T}_n]_{0r} & \equiv & \left[\mathbf{T}_n\right] _{0(N-r)} = 
  \left\{ \begin{array}{ccc}
    1 & \mbox{if} & \xi \leq p_r\\
    0 & \mbox{if} & \xi > p_r
  \end{array} 
  \right.,\\
  \left[\mathbf{T}_n\right] _{q w} & = & [\mathbf{T}_n]_{0( w - q \;\mbox{mod}N)},
\end{eqnarray*}
in which $r=0,1,\ldots,N'$, with $N'\nobreak\equiv\nobreak\frac{N-1}{2}$, $[\mathbf{T}_n]_{0N}\equiv[\mathbf{T}_n]_{00}$, $\xi$ is random variable, and $p_r$ is a parameter that determines the connectivity between sites 0 and $r$. We remark that the connectivity distribution of each site is uniquely defined by the parameter $p_r$. If  $\xi$ is $\delta-$correlated and with uniform probability distribution in $[0,1]$, then $p_r$ defines the coupling probability. Therefore, the probability of occurrence of a specific coupling  matrix  depends only on entries of the topology matrix:
\begin{equation}\label{EQ.def.pi}
	\pi_k = \prod_{r = 0}^{N'}\left(p_r\left[\mathbf{T}_k\right]_{0 r}+(1-p_r)(1-[\mathbf{T}_k]_{0r})\right).
\end{equation}
If the conditions (i) and (ii) are fulfilled, there are, at most, $K_N = 2^{\frac{N + 1}{2}}$ distinct matrices. For each $p_r=0$ or $p_r=1$, $K_N$ is reduced by a factor 2. The evolution of the network is then obtained by the selection of $\mathbf{G}_n$ from the set $\{\mathbf{G}_k\}_{k=1}^{K_N}$ according the probability $\pi_k$.

The translacional invariance of the lattice topology implies all sites share the same  connectivity degree $\kappa_n$ at the same instant $n$. Along the temporal evolution of the lattice, $\kappa_n$ follows a Poissonian  distribution, because we assume that the $p_r$'s are independent of each other. Different distributions can be obtained through a rescaling of the probabilities $\pi_k$ following their respective connectivities.

\section{\label{SEC.methods}Methods}

\subsection{Chaos synchronization}
We are interested in studying the CML (\ref{cml}) from the point of view of the synchronization manifold stability. Such manifold is defined by 
$$\mathcal{S} = \left\{\mathbf{y}\in\Omega, \mathbf{s}\in\omega:\mathbf{x}^{(m)} \equiv \mathbf{s} \forall m \right\},$$
that is, we are considering the {\it complete amplitude synchronization} -- all sites share the same state at same time. We say that a trajectory $\{\mathbf{y}_t \}_{t = 0}^n$ achieves $\mathcal{S}$ at time $n$ when, for any $\delta >0$, we observe $\hbox{dist} \left( \mathbf{y}_n, \mathcal{S} \right) < \delta.$ The mapping ${\bf F}$ keeps $\mathcal{S}$ invariant, and the dynamic in this subspace is determined by ${\bf g}$.

\subsection{Linear stability analysis}
The stability of the synchronization manifold can be determined through  the Lyapunov exponents' spectrum of $\mathcal{S}$, calculated for a typical trajectory in this manifold. The $m-$th Lyapunov exponent is defined by 
\begin{equation}\label{EQdeflyap}
  \lambda_m(\mathbf{s}_0) = \lim_{n\rightarrow\infty}\frac{1}{n}\ln\|\mathbf{D}\mathbf{F}^{(n)}(\mathbf{s}_0)\mathbf{e}_m\|,  
\end{equation}
in which $\mathbf{D}\equiv\frac{\partial}{\partial \mathbf{y}}$ is the $dN-$dimensional Jacobian matrix, $\mathbf{e}_m$ is a typical unitary vector in $\mathbb{E}^m \backslash \mathbb{E}^{m+1}$, and $\mathbb{E}^0=\Omega$ \cite{RMP.v57.p617.y1985}. Since the dynamics in $\mathcal{S}$ is ruled by the map $\mathbf{g}$, which is, by assumption, mixing, $\lambda_m (\mathbf{s}_0)$ do not depends on $\mathbf{s}_0$, apart from a set of null Lebesgue measure, and we have $\lambda_m (\mathbf{s}_0) \equiv \lambda_m$. The calculations yielded by (\ref{EQdeflyap}) give $\lambda_0 \geqslant \lambda_1 \geqslant \cdots \geqslant \lambda_{N d - 1}.$
The analysis of the stability of $\mathcal{S}$ is done by the examination of the greatest transversal Lyapunov exponent, denoted by $\lambda_{\perp}$, which is calculated in $\Omega\backslash\mathcal{S}$. In the case of networks whose topology varies over time, the scaling behaviour of the Hajnal diameter of the product of coupling matrices enables us to obtain $\lambda_{\perp}$ \cite{EJPB.v63.p399.y2008, SIAMJMA.v39.p1231.y2007}. 

Now, let us develop Eq. (\ref{EQdeflyap}) for typical trajectories embedded in $\mathcal{S}$. For the sake of simplicity, we shall consider $d=1$. Let $\mathbf{J}_n(\mathbf{y}_0)$ be the Jacobian matrix applied in (\ref{cml}), which is calculated over $\{\mathbf{y}_t\}_{t=0}^{n-1}$: 
$$\mathbf{J}_n (\mathbf{y}_0) = \prod_{t = 0}^{n - 1} \mathbf{G}_t \mathbf{D}\mathbf{f}(\mathbf{y}_t).$$
In $\mathcal{S}$ we have $\mathbf{D}\mathbf{f}(\mathbf{y}_t) = \frac{\partial g(s_t)}{\partial s}\mathbb{I}_N$, where  $\mathbb{I}_N$ is the identity  $N \times N$ matrix. \linelabel{line.rev10} Thus, the calculation of the Lyapunov exponents given by Eq. (\ref{EQdeflyap}) yields
\begin{equation}\label{EQlyapSS}
  \lambda_m = \lim_{n \rightarrow \infty} \frac{1}{n} \ln \|e^{\lambda_U n} \left(\prod_{t=0}^{n-1}\mathbf{G}_t\right)\mathbf{e}_m \|,
\end{equation}
in which $\lambda_U>0$ is the Lyapunov exponent of the map $g(s)$.

Once $\mathbf{G}_t$ is a matrix satisfying the conditions (i), (ii) and Eq. (\ref{conditionsA}), we can show that \footnote{\label{FN.matrices} Our results remain valid under less restrictive conditions with respect to the possible coupling matrices. Theorems in Refs. \cite{PPHajnal, PPWolfowitz} guarantee that the infinite product of  matrices that are indecomposable and aperiodic  converges to a matrix whose all rows are identical (matrix $\mathbf{H}$). A matrix is indecomposable if it contains no submatrix of order $s\times (N-s)$ composed of zeros, whatever $s> 0$. If there is a finite sequence whose product results in a strictly positive matrix, {\it i.e.} all matrix elements are positive, the infinite product of indecomposable matrices converges to the matrix $\mathbf{H}$. This result was used by Lu {\it et al.} in Refs. \cite{SIAMJMA.v39.p1231.y2007, EJPB.v63.p399.y2008} in the study of coupled map lattices in which the coupling is time-varying. They demonstrated the need for the union of all sets contains a spanning tree, that is, any pair of elements must be eventually coupled. We extend the applicability of expression (\ref{EQeffectivelyap}) for all Lyapunov exponents of the synchronization subspace, from which we could build an effective system (autonomous and deterministic). The restrictions that we imposed on the topology of the network allowed us to derive analytically an expression for the effective quantities, Eq. (\ref{EQdeftheta}).}

$$\lim_{n \rightarrow \infty}\prod_{t=0}^{n-1}\mathbf{G}_t = \mathbf{H},$$ with $[\mathbf{H}]_{q w} = [\mathbf{H}]_{0 w}$, for all $q$ and $w$. The values labeled by $[\mathbf{H}]_{0 w}$ depends on the observed coupling matrix chronological sequence, though it does not hold for the $\mathbf{H}$ form. $\mathbf{H}$ is a stochastic matrix, whose greatest eigenvalue is $\eta_0 \equiv 1$, which is associated to the eigenvector $\mathbf{h}_0 = \frac{1}{\sqrt{N}} \left[ \begin{array}{ccc} 1 & \cdots & 1 \end{array}\right]^{T}$. All the others eigenvalues, associated with the eigenvectors $\mathbf{h}_q$ $(q>0)$ are equal to zero, $\eta_q\equiv 0$.

\section{\label{SEC.results}Results and Discussion}

\subsection{The effective quantities}

In the limit $n \rightarrow \infty$, the Lyapunov vectors of $\mathcal{S}$, which are parallel to the eigendirections of the Jacobian matrix, are determined by $\mathbf{h}_q$. For this reason, the Lyapunov exponents are given by
$$\lambda_m = \lim_{n\rightarrow\infty}\frac{1}{n}\ln\|e^{\lambda_U n}\eta_m\|,$$
with $\mathbf{e}_m = \mathbf{h}_m$ in (\ref{EQlyapSS}) (the index of $\mathbf{h}_m$ are chosen to provide a non-increasing spectrum).

For $n \rightarrow \infty$ we have $e^{\lambda_U n}\rightarrow\infty$ and $\eta_m\rightarrow 0$ ($m>0$). If we write
\begin{equation}\label{EQeigenvaluedecay}
  \hat{\theta}_m = \left( \eta_m \right)^{1/n},
\end{equation}
we have
\begin{equation}\label{EQeffectivelyap}
  \hat{\lambda}_m = \lim_{n \rightarrow \infty} \frac{1}{n} \ln \| \left(e^{\lambda_U} \hat{\theta}_m \right)^n \| = \lambda_U + \ln | \hat{\theta}_m|.
\end{equation}

The quantities $\hat{\lambda}_m$ and $(e^{\lambda_U}\hat{\theta}_m)$ represent the $m$-th  effective exponent and Lyapunov number, respectively, and we call $\hat{\theta}_m$ the $m$-th effective coupling eigenvalue. For $n\gg 1$ we have $n\hat{\lambda}_m$, which describe the time-$n$ average behavior of the system. If $\eta_m$ approaches zero faster than the exponential, the limit  $n\rightarrow \infty$ in Eq. (\ref{EQeigenvaluedecay}) results $\hat{\theta}_m=0$, therefore, Eq. (\ref{EQeffectivelyap}) gives $\hat{\lambda}_m=-\infty$.

Expression (\ref{EQeigenvaluedecay}) gives a general and strong meaning to the effective coupling eigenvalues. The $\hat{\theta}_m$ ($m>0$) quantity is obtained by the decay rate to zero of the eigenvalues of the matrix $\mathbf{H}_n = \prod_{t=0}^{n-1}\mathbf{G}_t$. Only one condition was imposed, namely $\lim_{n\rightarrow\infty}\mathbf{H}_n = \mathbf{H}$. In other words, for any system in the form (\ref{cml}), in which all the possible coupling matrices satisfies (i), (ii) and Eq. (\ref{conditionsA}) (see \ref{AP.circulant}), the results (\ref{EQeigenvaluedecay}) and (\ref{EQeffectivelyap}) are formally valid. Thus, in general, our results suggest that the effective dynamics of a network whose topology is time-varying can be obtained by the analytical or numerical calculation   of $\hat{\theta}_m$.

Since the conditions (i) and (ii) imply all coupling matrices are circulant, we can write down the exact expression for (\ref{EQeigenvaluedecay}). This is possible because all circulant matrices share the same basis and, consequently, commute with each other. For this reason, (see \ref{AP.circulant})
\begin{equation}\label{EQdeftheta}
  \hat{\theta}_m = \prod_{k=1}^{K_N}\left(\Gamma_m^{(k)}\right)^{\pi_k},
\end{equation}
in which the productory extends over all possible coupling matrices, and $\Gamma_m^{(k)}$ is the $m$-th eigenvalue of the $k$-th coupling matrix. Here and in what follows, the $\Gamma^{(k)}_m$ (and the $\hat{\theta}_m$) are not indexed by their magnitudes, but according their respective eigenvectors whose components are $h_q^{(w)} \propto \exp\left(-(2\pi i)q w/N\right)$\cite{FTCIT.v2.p155.y2006}. Notice that, due symmetry, $\Gamma^{(k)}_m = \Gamma^{(k)}_ {N-m}$ and all the $\hat{\theta}_m$ with $m>0$ are degenerated. If for some $k$, $\Gamma_m^{(k)}=0$, then $\hat{\theta}_m=0$, which implies the system is superstable in the $m$-th direction. Strictly speaking, in this situation, any implementation of the system eventually results in superstability in the $m-$th direction. The synchronization manifold is transversely superstable if, and only if, $\hat{\theta}_m=0$ in all transversal directions, {\it i.e.} for every $m>0$. In average, the time required for a typical trajectory to experience the effects of superstability    is of the order of $\pi_k^{-1}$.

\subsection{The effective coupling}

Let $\mbox{diag}\{\hat{\theta}_m\}$ be a diagonal matrix formed by the effective coupling eigenvalues (\ref{EQdeftheta}). Let $\mathbf{P}$ be an unitary similarity transformation formed by the Lyapunov vectors of $\mathcal{S}$. Hence, we can make an effective {\it autonomous deterministic} system from the definition of an effective coupling matrix:
$$\hat{\mathbf{G}} = \mathbf{P}\mbox{diag}\{\hat{\theta}_m\}\mathbf{P}^\dagger.$$
From the point of view of the synchronization capacity we show that the systems constructed by $\{\mathbf{G}_n\}$ and $\hat{\mathbf{G}}$ in (\ref{cml}) are equivalent.

For the systems under consideration, we have $[\mathbf{P}]_{qw}=\exp(-(2\pi i)q w/N)/\sqrt{N}$, therefore
\begin{equation}\label{EQeffectivesystem}
  [\hat{\mathbf{G}}]_{q w} = \frac{1}{N}\left\{\hat{\theta}_0+2\sum_{s=1}^{N'}\hat{\theta}_s \cos\left(\frac{2\pi s(w-q)}{N}\right)\right\}.
\end{equation}

\begin{figure}[tb]
  \includegraphics[width=\shrinkfactor\hsize, clip]{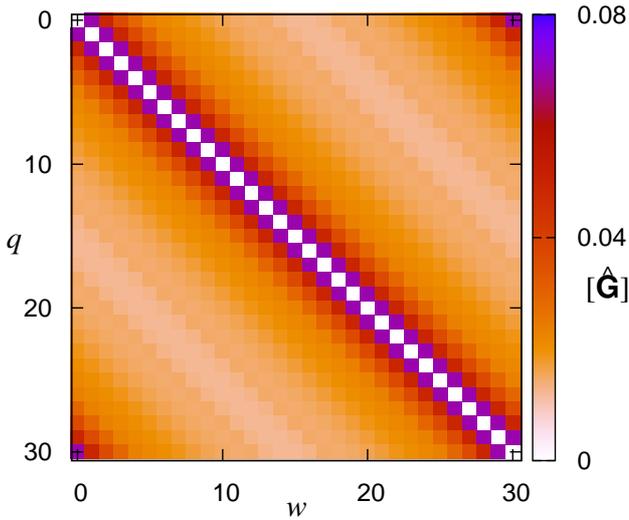}
  \caption{\label{FIGmatrix}(color online) Pictorial representation of $\hat{\mathbf{G}}$ elements, in which the color scale stands for the coupling intensity. Note the periodicity and translational invariance of the coupling. This matrix has the following parameters: $N = 31$, $\varepsilon = 0.7$ and $\alpha = 0.6$ (elements on the main diagonal are not shown).}
\end{figure}

Once $\cos(\cdot)$ is an even function with period $2\pi$ and $\hat{\theta}_0=1$, is straightforward that $\hat{\mathbf{G}}$ is a circulant, stochastic and symmetric matrix.  For this reason, the effective matrix satisfies all the suppositions about the network. An example is found in Figure \ref{FIGmatrix}.

In general, due to the explicit dependence of the Jacobian matrix on the points of the trajectory, we may formally demonstrate the equivalence between both systems only in the linear neighborhood of $\mathcal{S}$. When such dependence does not exist, the Lyapunov vectors of all points in phase space are identical and our results are globally valid in $\Omega$ \cite{PRE.v68.p045202.y2003, PhysicaA.v389.p5279.y2010}. From this, we conjecture that the equivalence between the two systems, for synchronization purposes, is generic. Our results are also valid for networks in which the topology varies periodically in time, replacing $\pi_k$ by the relative frequency associated with $k$-th topology.

\subsection{\label{SEC.example}Example}
We shall apply the above results to a system in which, at each time instant, two sites are coupled with probability  $p_r=r^{-\alpha}$; being $r$ the shortest distance between sites $0$ and $r$, $p_0\equiv 1$, and $\alpha\geq 0$ a parameter that quantifies the range of the coupling. We assume the same value of the coupling strength $\varepsilon$ for all connections. Therefore, the coupling is expressed by
\begin{equation}\label{EQlatticerealization}
  [\mathbf{G}_n(\mathbf{T}_n)]_{q w} = \Big([\mathbf{T}_n]_{qw}-\Big(1+(1-\varepsilon^{-1})\nu_n\Big)\delta_{qw}\Big)\frac{\varepsilon}{\nu_n},
\end{equation}
in which $\nu_n=2\sum_{r=1}^{N'}[\mathbf{T}_n]_{0r}$ is a normalization factor according to (\ref{conditionsA}). Since, by construction, $p_0=p_1=1$, the total number of different topologies for this system is $K_N = 2^{\frac{N-3}{2}}$.

This example describes a network with time-varying topology and interactions whose probability of coupling between two sites depends on the $r$ distance between them. This spatial dependence is recovered by the effective matrix $\hat{\mathbf{G}}$ in the form of the coupling intensities, as shown in Fig. \ref{FIGmatrix}.

\begin{figure}[tb]
  \includegraphics[width=\shrinkfactor\hsize, clip]{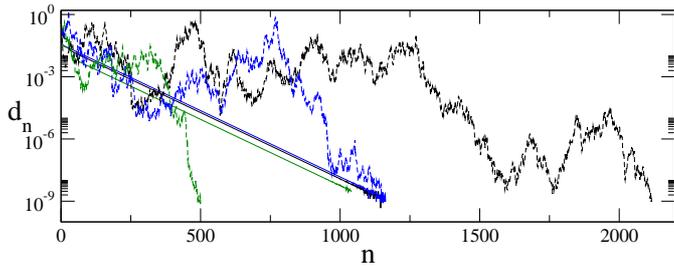}
  \caption{\label{FIGtemporal} (color online) The convergence to $\mathcal{S}$ of typical trajectories for the time-varying topologies (dashed lines) and for the effective system (solid lines), with $N=31$, $\varepsilon = 0.7$ and $\alpha= 0.6$.}
\end{figure}

We present an analysis on the synchronization of chaos in this system, as well as illustrations of the equivalence between the original system (\ref{cml}) and effective static system. In view of this, we consider the temporal evolution of the Euclidean distance between the points of the trajectory and $\mathcal{S}$, denoted by $d_n\equiv\mbox{dist}(\mathbf{y}_n,\mathcal{S})$. As a first examination, we consider the evolution of typical trajectories to $\mathcal{S}$ for a specific realization of this network. The results for the numerical simulations are presented in Figure \ref{FIGtemporal}. From the expressions (\ref{EQdeftheta}) and (\ref{EQeffectivelyap}) we determine a set of parameters $(N,\varepsilon,\alpha)$ for which $\mathcal{S}$ is transversely stable, in which we chose the tent map  $g(s) = 1 - 2|s-1/2|$ ($\lambda_U = \ln 2$) to rule the dynamics of each site. The rationale behind this choice is the monotonic convergence of the trajectories to the transverselly stable synchronized state, when static couplings are considered \cite{RFPandSESPtoappear2.y2010}. Thus, we can ensure that the irregular time evolution of $d_n$ in Fig. \ref{FIGtemporal} -- for typical trajectories outside the synchronization manifold -- is precisely due to the time-dependent coupling. For example, at a given instant $n$, it is possible that each site is only influenced by a few distant sites, which are in very different states, causing a departure from $\mathcal{S}$. On the other hand, the same mechanism may explain the decrease in the synchronization time when the states are very close together. Figure \ref{FIGtemporal} also shows the numerical simulations for the static performance of the system, in which we observe that the convergence is monotonic.

\begin{figure}[tb]
  \includegraphics[width=\shrinkfactor\hsize, clip]{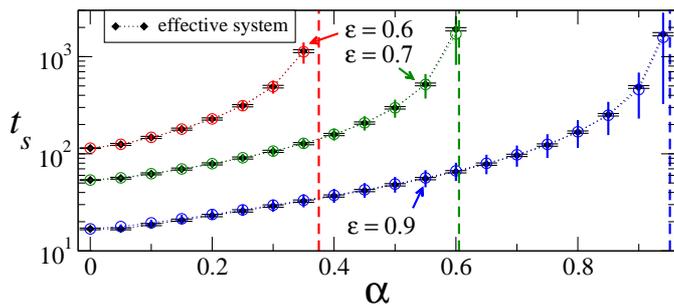}
  \caption{\label{FIGsynchtime} (color online) The dependence on the range of coupling probability of the average synchronization times, for the time-varying system (circles) and the effective static one (diamonds), and its standard deviation, given by vertical and horizontal bars, respectively, for $N = 31$. The vertical dashed lines are the critical values $\alpha_c$ for which $\mathcal{S}$ becomes transversaly unstable.}
\end{figure}

We must emphasize that typical trajectories of the system, whose topology is time-varying, present an average behavior that is determined by the equivalent static system, for any set of parameters -- including the lattice size -- inside the synchronization domain. This result is shown in Fig. \ref{FIGsynchtime} in which we analyze the average time required for synchronization of a set of typical trajectories. The synchronization time increases with the parameter $\alpha$ and diverges at its critical value, $\alpha_c(\varepsilon,N)$. For this parameter value $\mathcal{S}$ becomes transversely unstable. The $\alpha_c$ is determined by the equation (\ref{EQdeftheta}) when $|e^{\lambda_U}\hat{\theta}_\perp|=1$ \cite{PRE.v68.p045202.y2003}, being $\hat{\theta}_\perp = \max_{m>0}\{|\hat{\theta}_m|\}$. This critical value \footnote{This network can present superstability at $\varepsilon\nobreak=\nobreak(1\nobreak-\nobreak1/N)\nobreak\equiv\varepsilon_{ss}$ for any $\alpha<\infty$, because global coupling has probability $\pi_{gl}=(N'!)^{-\alpha}$. Typically, this superstability is not observed because the set of parameters, which is determined by $\varepsilon_{ss}$, has null measure in parameter space. For $\varepsilon=\varepsilon_{ss}\pm\delta\varepsilon$ the contribution to the expression (\ref{EQdeftheta}) of the eigenvalue associated with global coupling is given by $\delta\varepsilon^{(N'!)^{-\alpha}}$, {\it i.e.} it becomes evident only for very small networks. For this reason, even for $\varepsilon\approx\varepsilon_{ss}$, no manifestation of this phenomenon is observed.}   is indicated by the dashed lines in Fig. \ref{FIGsynchtime}. The transition to synchronization, which is characterized by changes in transversal stability of $\mathcal{S}$, occurs at the same point for both systems. In this figure, we also analyze the dispersion of synchronization times. The dispersion is computed by the standard deviation of several network realizations. This quantity is represented by vertical and horizontal bars in the case of systems with varying and static topology, respectively.

\section{\label{SEC.conclusions}Conclusions}
In conclusion, we studied here the synchronization of chaos in coupled map lattices whose topology varies semi-stochastically with time. Regarding the periodicity, the translational invariance of lattice topology, and that all sites are eventually coupled, we showed that a weighted average over all possible topologies determines the dynamics transverse to the synchronization manifold. This was quantified by the introduction of effective Lyapunov exponents, for which we derive an exact analytical expression. We also showed that, starting from such effective dynamical quantities, a system that is {\it deterministic and autonomous} -- which we call effective system -- can be constructed, so that its topology is static. We demonstrated the equivalence between the two systems by analyzing the stability of the synchronization manifold, and the average time needed for the typical trajectories of the network to reach it. Although the average behavior is the same for both systems, we showed that the time dependence of the coupling can cause great variability in time synchronization for the case of isolated observations.

We found an efficient method to numerically determine the effective quantities. This method is based on the eigenvalue decay rate of the coupling matrices' products, which may be a quick way to calculate the Lyapunov spectrum, because it usually involves a small number of products of matrices (compared to the number of possible topologies) \cite{Spectral.y2010} .

Finally, we expect that any network with time-varying topology can be represented by a circulant effective matrix, provided that the network is periodic and that just the probability of interaction (and not the topology) between elements is invariant over translations. Consequently, knowing the Lyapunov exponents that characterize the transversal dynamics to the synchronized state, we can find a good estimate for the effective coupling between the elements of the system under study. This is possible with the use of equation (\ref{EQeffectivesystem}) after inversion of equation (\ref{EQeffectivelyap}). Note that the Lyapunov exponents can be obtained from exact calculation, numerical simulations or the analysis of time series from experiments.

\section*{Acknowledgments}
This work has been made possible thanks to the partial financial support from the following Brazilian research agencies: CNPq, CAPES and Funda\c c\~ao Arauc\'aria. R.F.P. would like to acknowledge S. Sinha for presenting the Refs. \cite{PRE.v66.p016209.y2002} and \cite{PRE.v78.p066209.y2008} to him at the Dynamics Days South America 2010 coffee break.


\appendix
\section{\label{AP.circulant}Derivation of Eq. (\ref{EQdeftheta})}

 The restrictions on the topology and coupling restrictions allow an analytical derivation of the effective quantities. Since any matrix satisfying the conditions (i), (ii) and Eq. (\ref{conditionsA}) is circulant  \cite{FTCIT.v2.p155.y2006}. A matrix $N\times N$ is circulant  if its elements satisfy
$$ [\mathbf{G}]_{q,w} = [\mathbf{G}]_{0,(w-q \mod N)} \qquad \forall q,w=0,1,\cdots,N-1.$$

The $m-$th eigenvalue of a circulant matrix that is associated to the eigenvector with components $h_q^{(w)} \propto \exp\left(-(2\pi i)q w/N\right)$ is given by
\begin{equation*}
	\Gamma_m = \sum_{w=0}^{N-1}[\mathbf{G}]_{0w}e^{-(2\pi i)m w/N}.
\end{equation*}

If the matrix is circulant and symmetric, this expression reads

\begin{equation}
	\Gamma_m = [\mathbf{G}]_{00} + 2\sum_{w=1}^{N'}[\mathbf{G}]_{0w}\cos\left(\frac{2\pi m w}{N}\right),
\end{equation}
in which $N'\equiv (N-1)/2$.

Assume that $\mathbf{H}_n$ represents a specific sequence

\begin{equation}\label{EQ.ap.hn1}
	\mathbf{H}_n = \mathbf{G}^{(k_{n-1})}\mathbf{G}^{(k_{n-2})}\cdots\mathbf{G}^{(k_{1})}\mathbf{G}^{(k_{0})},
\end{equation}
being  $k_n$ a label that identifies a coupling matrix observed, at time $n$, in the set of the $K_N$ possible coupling matrices. Since all the matrices have the same basis, they commute among themselves and we can rewrite  the Eq.   (\ref{EQ.ap.hn1}) in the following way:

\begin{equation*}\label{EQ.ap.hn2}
	\mathbf{H}_n = \prod_{k=1}^{K_N}\left(\mathbf{G}^{(k)}\right)^{\nu_k},
\end{equation*}
in which $\nu_k$ represents how many times the $k$-th matrix $\mathbf{G}^{(k)}$ was observed, so $\sum_k \nu_k = n$. In the limit $n\to \infty$ we have $\nu_k = n\pi_k$, therefore

\begin{equation*}
\mathbf{P}^\dagger\mathbf{H}_n\mathbf{P} = \mathbf{P}^\dagger\prod_{k=1}^{K_N}\left(\mathbf{G}^{(k)}\right)^{\nu_k}\mathbf{P} = \prod_{k=1}^{k_N}\left(\mbox{diag}\{\Gamma^{(k)}_m\}\right)^{n\pi_k}.
\end{equation*}

Since $\mathbf{P}^\dagger\mathbf{H}_n\mathbf{P} = \mbox{diag}\{\eta_m\}$,  Eq. (\ref{EQdeftheta}) is obtained by taking the $n$-th root of expression above.

\bibliographystyle{elsarticle-num}
\bibliography{biblio.all.bib}

\end{document}